\documentclass[%
 reprint,
superscriptaddress,
 amsmath,amssymb,
 aps,
pre,
]{revtex4-2}

\usepackage{graphicx}
\usepackage{dcolumn}
\usepackage{bm}
\usepackage{xcolor}

\begin{document}

\preprint{APS/123-QED}

\title{
Validation and parameterization of a novel physics-constrained neural dynamics model applied to turbulent fluid flow
}

%

\author{Varun Shankar}
\affiliation{Carnegie Mellon University}

\author{Gavin D. Portwood}
\affiliation{Los Alamos National Laboratory}
\affiliation{Lawrence Livermore National Laboratory}

\author{Arvind T. Mohan}
\affiliation{Los Alamos National Laboratory}

\author{Peetak P. Mitra}
\affiliation{University of Massachusetts Amherst}
\affiliation{Palo Alto Research Center}

\author{Dilip Krishnamurthy}
\affiliation{Carnegie Mellon University}

\author{Christopher Rackauckas }
\affiliation{Massachusetts Institute of Technology}

\author{Lucas A. Wilson}
\affiliation{Dell Technologies}

\author{David P. Schmidt}
\affiliation{University of Massachusetts Amherst}

\author{Venkatasubramanian Viswanathan}
\affiliation{Carnegie Mellon University}
\email{venkvis@cmu.edu}

\date{\today}

\begin{abstract}
In fluid physics, data-driven models to enhance or accelerate solution methods are becoming increasingly popular for many application domains, such as alternatives to turbulence closures, system surrogates, or for new physics discovery. In the context of reduced order models of high-dimensional time-dependent fluid systems, machine learning methods grant the benefit of automated learning from data, but the burden of a model lies on its reduced-order representation of both the fluid state and physical dynamics. 
In this work, we build a physics-constrained, data-driven reduced order model for the Navier-Stokes equations to approximate spatio-temporal turbulent fluid dynamics. The model design choices mimic numerical and physical constraints by, for example, implicitly enforcing the incompressibility constraint and utilizing continuous Neural Ordinary Differential Equations for tracking the evolution of the differential equation. We demonstrate this technique on three-dimensional, moderate Reynolds number turbulent fluid flow. In assessing the statistical quality and characteristics of the machine-learned model through rigorous diagnostic tests, we find that our model is capable of reconstructing the dynamics of the flow over large integral timescales, favoring accuracy at the larger length scales. More significantly, comprehensive diagnostics suggest that physically-interpretable model parameters, corresponding to the representations of the fluid state and dynamics, have attributable and quantifiable impact on the quality of the model predictions and computational complexity.
\end{abstract}

\maketitle


\section{Introduction}
Model reduction is commonly employed by scientists and engineers to obtain
solutions to complex turbulent fluid systems which may otherwise be
prohibitively complex to simulate by their physical governing equations. Indeed
various reduced order modeling approaches have been proposed for turbulent fluid systems
which have the potential to aid in applications such as simulation-based design
optimisation or statistical inverse modeling.  Recently,
reduced order modeling of turbulent fluid systems by various deep learning
techniques has shown strong promise in accurately modeling various aspects of
fluid motion while being computationally efficient compared to their
first-principles counterparts.  While modeling approaches which demonstrate
favorable accuracy with respect to physical diagnostics are encouraging, we
believe more rigorous and interpretable parameterization of approximation
error is essential for model certification in predictive science.

The understanding of errors in machine-learned reduced order models is complicated by the lack of formal
approximation theory for many classes of data driven models, in particular our interest in deep feed-forward and convolutional
neural networks.  Combined with the tendency for surrogate models to handle
disparate phenomena with so-called deep ``end-to-end'' architectures, it is
frequently unclear how to attribute model error in deep neural networks with respect to model architectures.

However, various efforts have improved the state of model error
interpretability. Firstly, there are various approaches in which physical constraints can be imposed on ML models \citep{Brunton2020}. For instance, the development of loss functions which are informed by governing equations in various manifestations of physics-informed neural networks (PINNs)  have demonstrated effective surrogate models which approximate physical solutions \citep{PINN_2019}. Network architectures which implicitly preserve hard physical constraints have shown promise in improving model accuracy by enforcing strictly physically-realizable model outputs in some respects \citep{Wang_2020,mohan2020embedding}.

Beyond enforcing physical constraints, various approaches have been suggested using convolutional neural networks (CNNs) \cite{Thuerey_2020,Wang_2020,Mohan2019CompressedCL,belbuteperes2020combining}, recurrent neural networks (RNNs) \cite{Brunton2020} or long short-term memory (LSTM) \cite{Mohan2019CompressedCL,Vlachas2018} architectures. These models typically encode spatial information from the field using CNNs and use autoregressive predictions or RNN modules to address the temporal dependency. Beyond reduced-order modeling, other machine learning methodologies have been proposed with physics-informed components, such as methods to relax grid-resolution requirements with learned interpolations or mesh coarsening corrections for the simulation \cite{Kochkov2021}, accelerate intermediate solver steps \cite{tompson17}, and improve predictive capacity with hybrid CFD and ML-correcting algorithms \cite{portwood2021interpreting, belbuteperes2020combining, watt2021correcting}.

However, the adoption of various manifestations of differential
equation-based deep learning models, such as NeuralODEs (NODEs), have  effectively demonstrated the shift of modeling
burden to numerically-formal time integrators \cite{Chen_2019}. NeuralODEs have previously been used for forecasting limited turbulence prognostics and system states \cite{Portwood_2019, mohan2021learning}. More importantly, this framework establishes a principled way to trade-off accuracy and depth, and offers a step towards bridging the fields of numerical methods and machine learning. 

Leveraging NeuralODE and PINN design concepts in surrogate models, deep learning architectures may be developed to mimic formal numerical solution techniques such that neural network subcomponents are meaningful and intuitive. Therefore, we utilize intuitive design concepts in the development of a reduced-order model for three dimensional solutions to the incompressible Navier-Stokes equations as detailed in section 2 of this manuscript. In addition to enforcing hard physical constraints, we carefully design our deep network architecture to explicitly treat two discrete modeling demands in distinct subnetworks: reduction of physical states and reduction of system dynamics.  By parameterizing the architecture the two subnetworks, we demonstrate an effective modeling methodology with physically-relevant diagnostics. More importantly, we investigate the behaviour of model error as a function of individual subnetwork complexity as a means to systematically demonstrate trade-offs between modeling error and computational complexity.

\section{Problem statement}
We consider incompressible homogeneous isotropic turbulence (HIT) as a fundamental model flow that features sufficient complexity to evaluate our proposed modeling methodology introduced above.  In addition to exhibiting strong nonlinearity in the parameter regimes we consider, the multiscale structure of HIT makes the reduced-order representation of flow states a non-trivial task.  Such modeling challenges are compounded by complexity in the governing equation, which induce non-local phenomena due to the incompressibility constraint. 
The equations of motion we consider are given by the momentum equations and mass balance:
\begin{equation}
\frac{\partial \mathbf{u}}{\partial t}+(\mathbf{u}\cdot\nabla)\mathbf{u}=\nu\nabla^2\mathbf{u}-\frac{1}{\rho}\nabla p+\mathbf{f} ;\quad \nabla\cdot\mathbf{u}=0
\label{eq:ns}
\end{equation}
where $\mathbf{u}$ is the velocity vector, defined $\{ u,v,w\}$, $p$ is the pressure, $\rho$ is the density, $\nu$ is kinematic viscosity and $\mathbf{f}$ is a dynamic large-scale forcing function which enforces statistical stationary in the turbulent kinetic energy \citep[][]{overholt98}, defined
\begin{equation}
e\equiv\frac{1}{2}(\langle u^2\rangle+\langle v^2\rangle+\langle w^2\rangle) \quad .
\label{eq:tke}
\end{equation}
We design our flow system with spatio-temporal homogeneity, by enforcing statistically stationary kinetic energy and periodic boundary conditions, to remove correlations with space or time which have the potential to be trivially memorized by our neural networks. 

\subsection{Modeling objectives}
Due to the combined multi-scale and non-linear behaviour of the Navier-Stokes equations, discretized direct numerical solutions to \eqref{eq:ns} are subject to unfavorable scaling of computational complexity with respect to the Reynolds number, where the number of grid points can be shown to scale with $Re^{9/4}$ \citep{pope}.
To improve computation tractability, much work has been done in reduced order modeling of dynamical systems. Proper orthogonal decomposition, often used with fluid flows, takes a statistical approach, projecting the data onto a linear subspace comprised of dominant modes \cite{Weiss_2019}. The idea of a latent space representation is common in many fields, including computer vision. Convolutional autoencoders (ConvAE) have been used for tasks like image compression \cite{Toderici2016,Rippel2017} or feature extraction \cite{Masci2011}, both of which closely parallel our goal of a rich reduced representation of the turbulent flow field. Data-driven ROMs are gaining traction in fluid applications as well \cite{Erichson2019PhysicsinformedAF, Gonzalez2018DeepCR}. Our approach takes advantage of this technique to reduce instantaneous velocity snapshots into a much smaller latent space. The dynamics of the system can then be performed on the latent representation, easing computation. Using a latent space in conjunction with NeuralODEs has been previously used successfully for modeling many dynamical systems \cite{latentNODE}.

The aim of the surrogate model is to track the spatiotemporal dynamics of the system in order to obtain approximate solutions.  Given an initial condition for the velocity field, the model should estimate the solution to the initial value problem (IVP) by forecasting the velocity field in time over a finite horizon. Modeling the dynamics of the spatially discretized velocity field as a differential equation is one method of achieving temporal evolution. This construction enables use of the NODE framework, which models the dynamics of a state variable through a neural network. The network assumes the responsibility of learning the differential dynamics of the state, but the IVP can be solved with any number of formalized ODE integrators.

The challenge with combining ML models with formal numerical methods is the computational cost of backpropagation. As an alternative, adjoint-state methods has been a well-established numerical method for computing gradients of some functional. Chen et al. proposed using this method for efficiently calculating the gradients of a parameterized differential equation by solving the adjoint problem \cite{Chen_2019}. This dual form, which is well-defined for many classes of differential equations, can be solved using traditional numerical methods as in the forward computation. The resulting parameter sensitivities can then be used to optimize the cost function, in this case, the reconstruction loss. In the context of deep-learning, the adjoint approach allows for a myriad of new architectures that permit differential equation solves to be easily integrated within the network structure, effectively instantiating a continuous time, infinite-depth network. Beyond the memory-efficient adjoint-optimization, NODEs permit arbitrary sampling in time, flexibility in the ODE integrator, among other benefits. The ability to adapt neural networks to include differential equations has far-reaching implications for modeling physical phenomena, which are often strictly described by ordinary or partial differential equations.

In order to incorporate the spatial information necessary to accurately approximate the true function, we use a convolutional neural network (CNN) to learn systems dynamics within the NODE model. 
This provides two benefits: (1) the memory costs of the network are drastically reduced as the number of parameters is much smaller than a fully-connected network. This is due to (2) the network uses a shared local filter in each layer that limits the message passing to a constrained receptive field around each output point. In a finite-layer CNN, the output at a location is determined from input values in a local neighborhood. The size of this neighborhood, or receptive field, is influenced by the number of layers in the network and the kernel sizes of the CNN layers.

The convolutional NODE formulation is a form of inductive, or learning bias imposed on the model. Although the neural network allows for much flexibility to be fit to the true data, it is still limited by the formulation, which enforces that the derivative of the velocity field with respect to time is a function of spatially local quantities, consistent with our knowledge of the physical system. Inductive biases are especially advantageous for generalization beyond the training set, as they encompass the assumptions a researcher makes about unseen data. In scientific machine learning applications, physics-based learning biases are a natural way to restrict models to prevent overfitting and produce outputs that more closely align with what is known about the system.


\section{Machine learning approach}

\subsection{Neural network architectures}

The encoder, denoted by $\mathcal{E}$, is a trainable network that transforms the input velocity field $\mathbf{u}_0$ to a lower dimensional latent space $\mathbf{z}_0$, given by
\begin{equation}
\mathcal{E}(\mathbf{u}_0) = \mathbf{z}_0 \quad .
\end{equation}
Compression is achieved with the use of strided convolutional layers in the encoder to reduce the spatial resolution in the latent space. These strided layers use the same convolutional spatial filtering technique to transform the input but coarsen the data by subsampling every other point (for a stride 2 layer). The size of the latent space varies in spatial resolution, corresponding to the number of strided, or coarsening layers, and latent channels, greater than the 3 input velocity channels. The encoder/decoder subnetworks are parameterized by the compression ratio, $z$, defined as the size of the original space relative to the latent space:
\begin{equation}
z=\frac{3 \cdot 64^3}{C_z N_z^3} \quad ,
\end{equation}
where $N_z$ is the number of points in each latent spatial dimension and $C_z$ is the number of latent channels. For reference with respect to physical quantities, the receptive field size of the encoder outputs are therefore equivalent to the grid spacing at $z=1$ and the domain size at $z=64$.
 
\begin{figure}[h]

  \centering

  \includegraphics[width=0.9\linewidth]{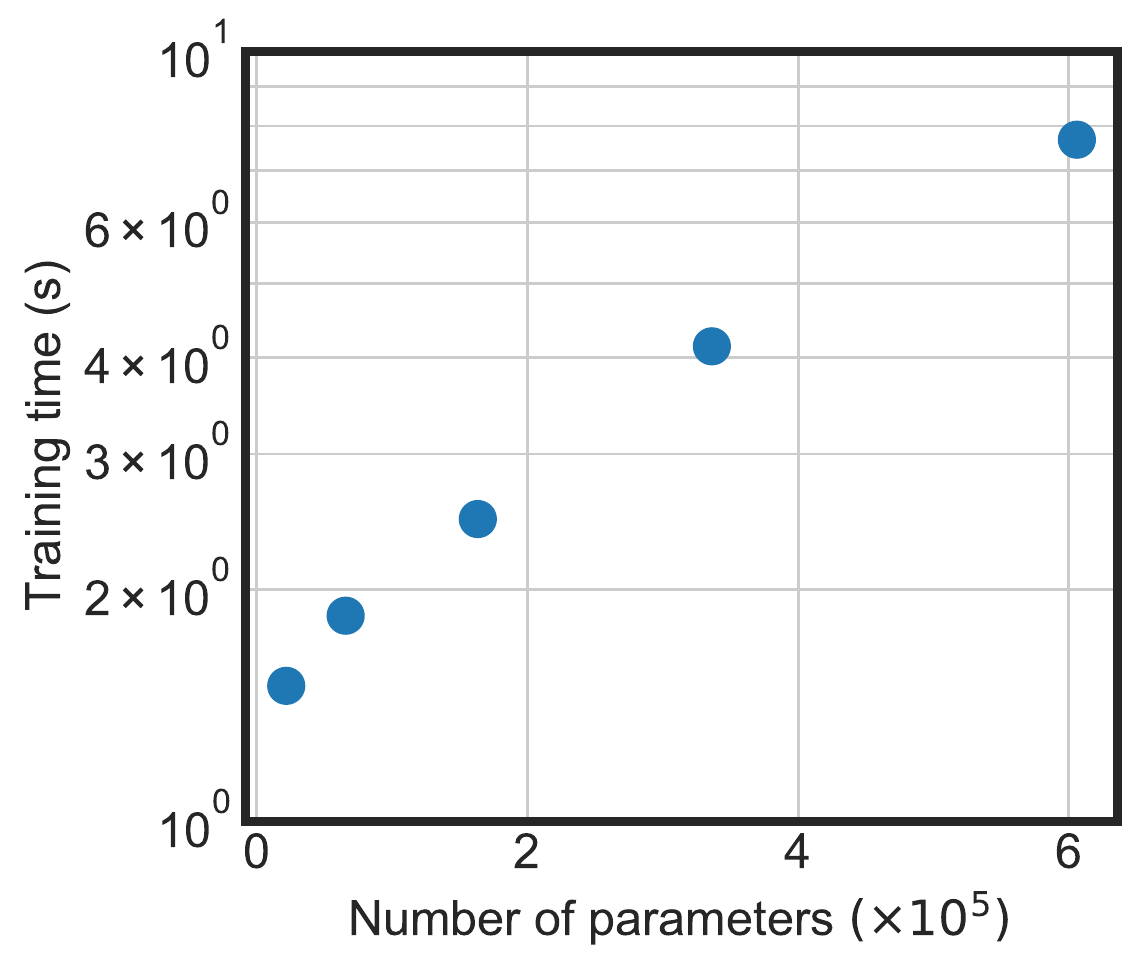}

  \caption{\small{Training time in seconds per sample per epoch for 5 kernel sizes (3, 5, 7, 9, 11), shown on a log scale.}} 
  \label{fig:training_time}

\end{figure}

 The initial condition in the latent space $\mathbf{z}_0$ is used to solve the ODE parameterized by the network weights. The system is defined by
\begin{equation}
\mathrm{\frac{d \mathbf{z}}{d t}=g_\theta(\mathbf{z}), \quad \mathbf{z}(0) = \mathbf{z}_0 \quad} ,
\end{equation}
where $g_{\theta}$ represents the dynamical convolutional subnetwork that learns the continuous dynamics of the latent state. The dynamical subnetwork consists of 3 convolutional layers and is parameterized by the kernel size of the layers. The kernel size of the NODE has a direct impact on the receptive field of the network and thus the sparsity of the ODE system, and in effect gives an indication of the length scale associated with the dynamics. Larger kernel sizes can learn higher order approximations of the spatial information, but this comes at the disadvantage of computational costs scaling as $O(k^3)$, where $k$ is the kernel size, for this 3-dimensional application. Figure \ref{fig:training_time} illustrates how the cost of a training iteration scales with kernel size and therefore, the number of network parameters. The points indicate models with $z=6$ and odd kernel sizes from 3 to 11. The scaling is consistent with using higher order numerical schemes. 

\begin{figure*}
  \centering
  \includegraphics[width=0.8\linewidth]{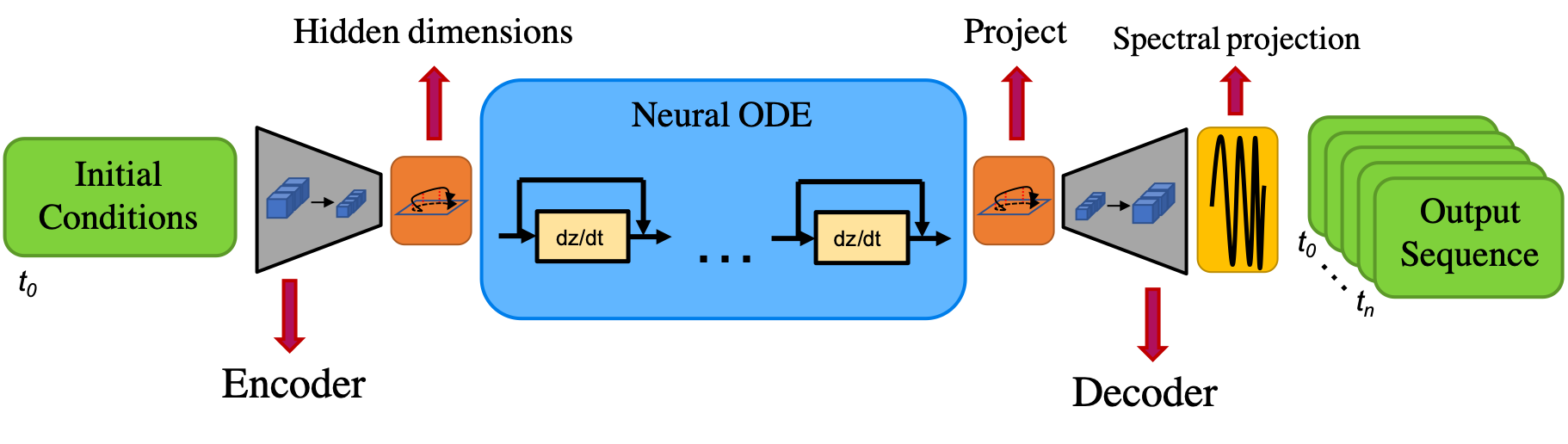}

  \caption{\small{Schematic of the overall model architecture. (Encoder): initial conditions are given and encoded into a latent space with a CNN. (Hidden dimensions): augmented channels are concatenated to increase dimensionality of the latent space. (NeuralODE): the latent dynamics are forecasted through the NeuralODE approximator. (Project): augmented channels are removed by projecting down into the original latent space. (Decoder): the sequence is decoded via another CNN. (Spectral projection): the divergence-free condition is enforced through the last layer.}
  }
  \label{fig:arch}

\end{figure*}

Although $\mathbf{z}(t)$ is a continuous function, and evaluated as such, the solution is saved at discrete time points corresponding to the training and test set. The decoder, $\mathcal{D}$, is another trainable network with mirrored architecture to that of the encoder, which decodes the latent sequence back to the original velocity space as
\begin{equation}
\mathcal{D}(\{\mathbf{z}_0,\mathbf{z}_1,...,\mathbf{z}_n\}) = \{\mathbf{\hat{u}}_0,\mathbf{\hat{u}}_1,...,\mathbf{\hat{u}}_n\} \quad .
\end{equation}
The network architecture is visualized in figure \ref{fig:arch}.


In order to build a more detailed understanding of the model, we run a series of sensitivity studies to determine the effect of the two subnetwork parameterizations on model accuracy.  In our experiments, we found that certain hyperparameters such as the size of the hidden layers did not have a significant impact on the results. However, the two hyperparameters with strong connections to traditional numerical methods, the compression ratio and the kernel size of the NeuralODE, were found to be a good indicator of model performance at varying scales. 

In our a-posteriori analysis, we test 13 different models: 4 compression ratios (6, 12, 24, and 48) and 3 kernel sizes (3, 5, and 7), shown in table \ref{tab:model_tab}, to determine their impact. We also examined a ``deep'' model, which used a compression ratio of 6 and a 9-layer CNN with kernel sizes of 3 in the NeuralODE, as opposed to the 3-layer networks in the other models. We will reference these models with the key-names $\mathbf{k}\alpha \mathbf{z} \beta$, where $\mathbf{k}$ and $\mathbf{z}$ represent the kernel size and compression ratio respectively, and $\alpha$ is the kernel size and $\beta$ is the compression ratio. The ``deep'' model is indicated with $deep$. The $deep$ model investigates whether the larger depth of the model, which increases the overall NeuralODE receptive field, can accommodate for the smaller kernels in each layer. The receptive field of the 9-layer $deep$ network is equivalent to the 3-layer $k7z6$ network.


\renewcommand{\arraystretch}{1.3}
\setlength{\tabcolsep}{5pt}
\begin{table}[]
\begin{tabular}{c c c c c}
                          & k=3            & k=5            & k=7            & \multicolumn{1}{l}{}              \\ \cline{2-5} 
\multicolumn{1}{l|}{z=48} & \textit{k3z48} & \textit{k5z48} & \textit{k7z48} & -                                 \\
\multicolumn{1}{l|}{z=24} & \textit{k3z24} & \textit{k5z24} & \textit{k7z24} & -                                 \\
\multicolumn{1}{l|}{z=12} & \textit{k3z12} & \textit{k5z12} & \textit{k7z12} & -                                 \\
\multicolumn{1}{l|}{z=6}  & \textit{k3z6}  & \textit{k5z6}  & \textit{k7z6}  & \multicolumn{1}{l}{\textit{deep}}
\end{tabular}
\caption{\small{Outline of the models tested in our studies. $k$ and $z$ denote the kernel size and compression ratio respectively. $deep$ denotes a 9-layer network with kernel sizes of 3.}} 
  \label{tab:model_tab}
\end{table}

\subsection{Physically-inspired model constraints}

We incorporate a few architectural elements to improve prediction and enforce physical constraints. As the problem includes periodic boundary conditions, this constraint is imposed by padding inputs to all convolutions circularly.
We also employ the augmented NeuralODE approach  to the latent dynamics, which is capable of learning a richer set of dynamics by representing the ODE on a higher dimensional space than the latent space input \cite{dupont2019augmented}. This is accomplished by concatenating channels of zeros to the output of the encoder, with the predicted results projected back into the original latent space. We use 2 augmented channels in our model, selected through parameter sweeps. 

Lastly, we apply a spectral projection operation to the final model output to enforce the divergence-free field condition required for a constant density fluid \cite{jiang2020enforcing}. The divergence-free velocity field comes from the mass balance in \eqref{eq:ns}. In this layer formulation, the input $\mathbf{u}'$ is transformed into Fourier space
\begin{equation}
    \mathbf{U}'=\mathcal{F}(\mathbf{u}') \quad ,
\end{equation}
where the divergence operator can be applied linearly. With this linear constraint, we can construct a quadratic optimization problem to minimize the squared difference between the input and output in Fourier space: 
\begin{equation}
    \begin{aligned}
    \min_{\mathbf{\hat{U}}} \quad & \frac{1}{2}(\mathbf{U}'-\mathbf{\hat{U}})^T(\mathbf{U}'-\mathbf{\hat{U}}) = \\
    \min_{\mathbf{\hat{U}}} \quad & \frac{1}{2}\mathbf{\hat{U}}^T\mathbf{I}\mathbf{\hat{U}} -
    \mathbf{U}'^T\mathbf{\hat{U}}\\
    \textrm{s.t.} \quad & \mathbf{A}\mathbf{\hat{U}} = \mathbf{0} \quad ,
    \end{aligned}
    \label{eq:div_opt}
\end{equation}
where $\mathbf{A}$ is the linear divergence operator. The solution can be obtained directly:
\begin{equation}
    \mathbf{\hat{U}} = \mathbf{U}'-\frac{\mathbf{k}\cdot \mathbf{U}'}{\mathbf{k}\cdot\mathbf{k}}\mathbf{k} \quad ,
\end{equation}
where $\mathbf{k}$ is a Fourier wavenumber vector. The final output comes from taking the inverse Fourier transform of the optimum,
\begin{equation}
    \mathbf{\hat{u}}=\mathcal{F}^{-1}(\mathbf{\hat{U}}) \quad .
\end{equation}

Each operation in the layer is fully differentiable and thus backpropagation through the layers with automatic differentiation is accomplished by the software without manual intervention.

\subsection{Data and Training}
We obtain solutions to \eqref{eq:ns} for a triply-periodic domain with direct numerical simulation (DNS) using standard Fourier psuedo-spectral method \citep[c.f.][]{orszag1972numerical}.
    The system is forced at low-wavenumbers to keep the total energy in the system constant \citep{overholt98}. The resulting data are statistically stationary with approximately stationary energy spectra. Our training, validation and test datasets span 100 integral time scales $\mathrm{\tau}\equiv e / \epsilon$, where $\epsilon$ is the turbulent kinetic energy dissipation rate. The turbulent Reynolds number, $\mathrm{Re_T \equiv e^2/(\nu \epsilon)}$ is approximately 380. The solutions are discretized with 64 collocation points in each direction, for a total of $64^3$ points and de-aliased with a standard two-thirds Fourier truncation filter.  We collect velocity fields every $\tau/100$, which are used to train the neural network.
    
    While the algorithm accepts a single instantaneous velocity field as input, during training, the trajectory is regressed with sampled temporal DNS snapshots from the forecasting window. The dynamics are learned from gradient-based optimization of the predicted sequence sampled at the same temporal rate as the DNS. From the NODE formulation, the prediction horizon can be adjusted on-the-fly, during training or validation. We use the normalized mean squared error (MSE) as the cost function for optimization. For this problem, we normalize the loss with the average energy in the flow. The loss can be written as:
    \begin{equation}
    \mathrm{L(\mathbf{u},\mathbf{\hat{u}})=\frac{\sum_{t=0}^{n} \langle \lVert \mathbf{u}_t - \mathbf{\hat{u}}_t \rVert^2 \rangle}{ e} \quad} ,
    \end{equation}
    where $\mathbf{u}_t$ and $\mathbf{\hat{u}}_t$ are the true and predicted velocity fields at time step $\mathrm{t}$, and $\mathrm{e}$ is given by eq.\ \eqref{eq:tke} and averaged over time.
    
    All models were trained with the ADAM optimizer using a decaying learning rate from $10^{-3}$ - $10^{-5}$. Additionally, models were trained by progressively increasing the prediction horizon from $\mathrm{\tau^*=0.25}$ to $\mathrm{\tau^*=1.0}$, where $\mathrm{\tau^*=t/\tau}$. Each model was trained for a total of 3500 epochs, 2000 at $\mathrm{\tau^*=0.25}$, 1000 at $\mathrm{\tau^*=0.5}$, 200 at $\mathrm{\tau^*=0.75}$, and 300 at $\mathrm{\tau^*=1.0}$. Based on a-posteriori analysis of the turbulence prognostics described in the subsequent section, this was determined to be sufficient for convergence. The training set spanned $10\mathrm{\tau}$, validation another $10\mathrm{\tau}$, and the test data encompassed the rest of the $80\mathrm{\tau}$ in the dataset. Our model is implemented in Julia using the \emph{DifferentialEquations.jl} and \emph{Flux.jl} packages \cite{rackauckas2017differentialequations}. We use the Tsitouras 5/4 Runge-Kutta method \cite{Tsitouras2011} for explicit time integration of the ODE dynamics. Model training was accomplished with NVIDIA V100 GPUs with a memory of 32GBs. GPU memory was a critical resource due to the large size of the dataset.


\section{Results}

We analyze the predictive capabilities of our model by examining a series of turbulent statistical metrics and comparing them to those of the DNS dataset. While standard statistical measures like local RMSE are often employed in machine learning applications to judge performance of models, and indeed we use the local MSE in our loss function, these values are limited in their physical insight and ultimately only offer a holistic qualitative measure of accuracy compared to the true data. 
In this section, we analyze data obtained from the DNS solutions at the suite of models described in the previous section, evaluated for one large eddy turnover time from an initial condition unseen during training.

\begin{figure}
  \centering
  \includegraphics[width=0.9\linewidth]{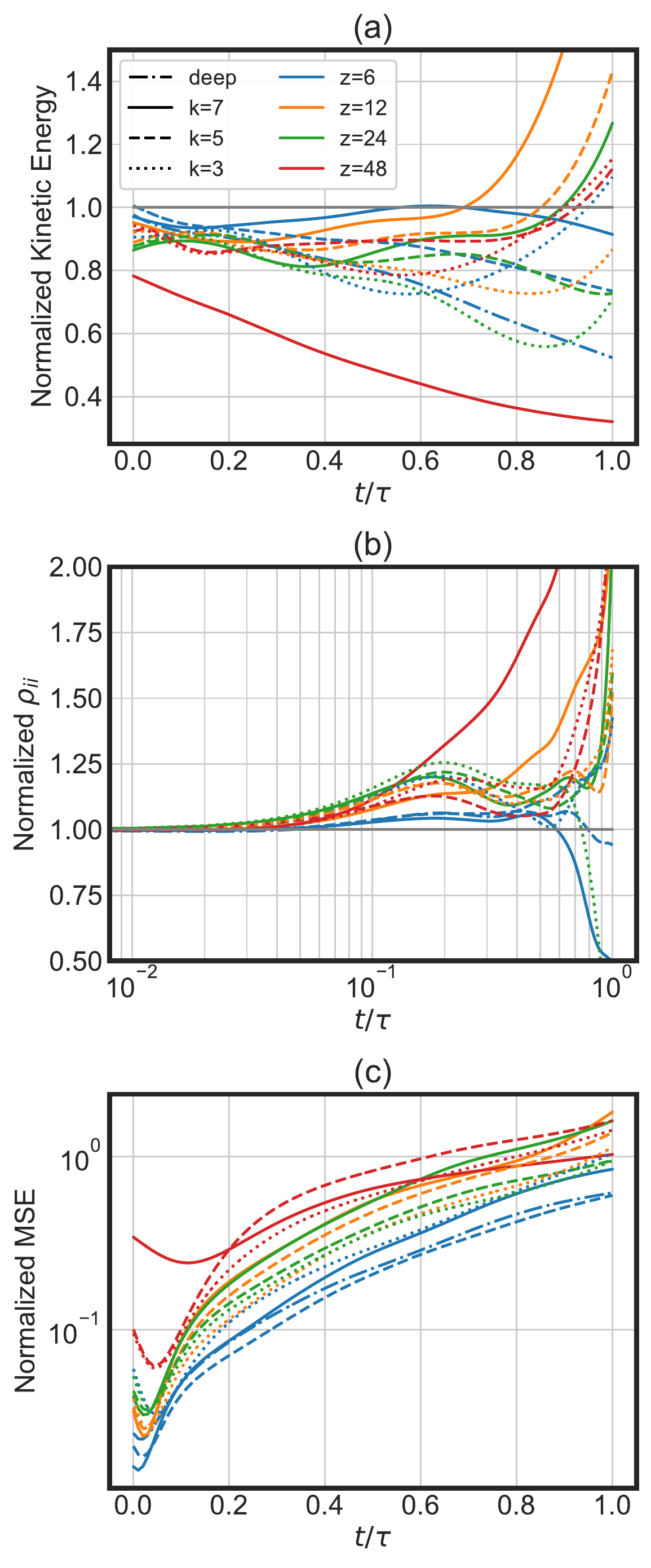}
  \caption{\small{We show three statistics that vary with time. The prediction window is normalized to the integral timescale of the flow. (a) turbulent kinetic energy in the flow. The predicted flow loses some energy but remains within reasonable bounds to the DNS data throughout. (b) velocity autocorrelation function on a log-log plot. (c) the MSE of the prediction normalized to the average energy in the flow.}} 
  \label{fig:temporal}
\end{figure}

\begin{figure}
  \centering
  \includegraphics[width=\linewidth]{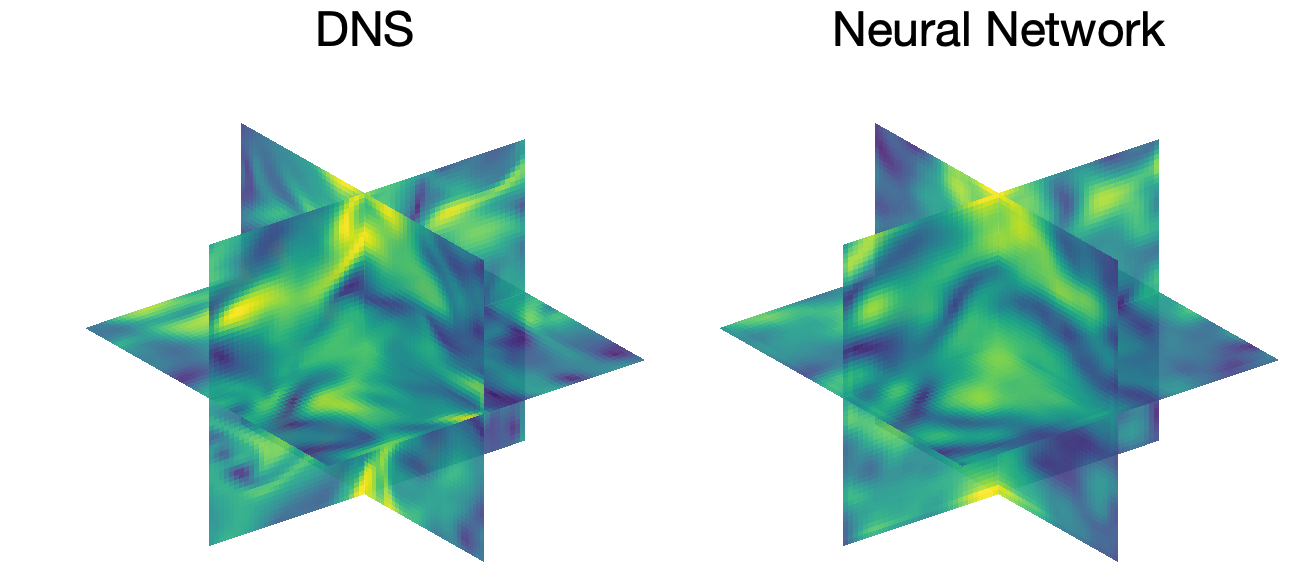}
  \caption{\small{Qualitative visual comparison of the ground truth DNS data and the output of the $k7z6$ network. Shown here is a 3D snapshot of the velocity magnitude at $\tau *=0.5$. Large scale structures appear to be preserved while small scale fluctuations are filtered out.}}
  \label{fig:qual}
\end{figure}

Figure \ref{fig:temporal}a depicts the turbulent kinetic energy in the flow, given by eq. \ref{eq:tke}, as a function of $\tau^*$ for each of the 13 models in table \ref{tab:model_tab}. 
We normalize each of the trajectories by the kinetic energy of the DNS, which deviates by less than one percent in this time window. 
At $\tau^*=0$, we observe each model to systematically underpredict the kinetic energy density at the initial condition.  We expect that our encoder subnetwork, which reduces the representation of the velocity field state as embodied by the compression ratio $z$, to resolve a subset of turbulent fluctuations. 
Indeed, at $\tau^*=0$ we observe the kinetic energy resolved by the velocity field reduction to decrease as the $z$ increases. Such phenomenology can be described by the encoder's exposure to larger and more energetic scales as $z$, and therefore the receptive field size, increases.
Such phenomenology is observed in figure \ref{fig:qual}, which provides a qualitative visual comparison of the output of the network with the DNS ground truth midway through the prediction window at $\tau^*=0.5$. Generally, large-scale structures appear to be preserved, while smaller scale fluctuations are filtered out by the model. 
At later times in figure \ref{fig:temporal}a, with the exception of $k7z48$, which appears to be an outlier, the models tend to slightly underpredict the flow energy in the initial phase before diverging more significantly after $\tau^*=0.5$. 
Solutions obtained from many models go unstable by $\tau^*=1$.
Models $k7z6$, $k5z6$, and $deep$ are the few that disregard this trend, which appear to have stable, perhaps overdissipative, solutions by 
$\tau^*=1$. Clearly, $k7z6$ tracks the DNS kinetic energy the closest.

\begin{figure}
  \centering
  \includegraphics[width=0.9\linewidth]{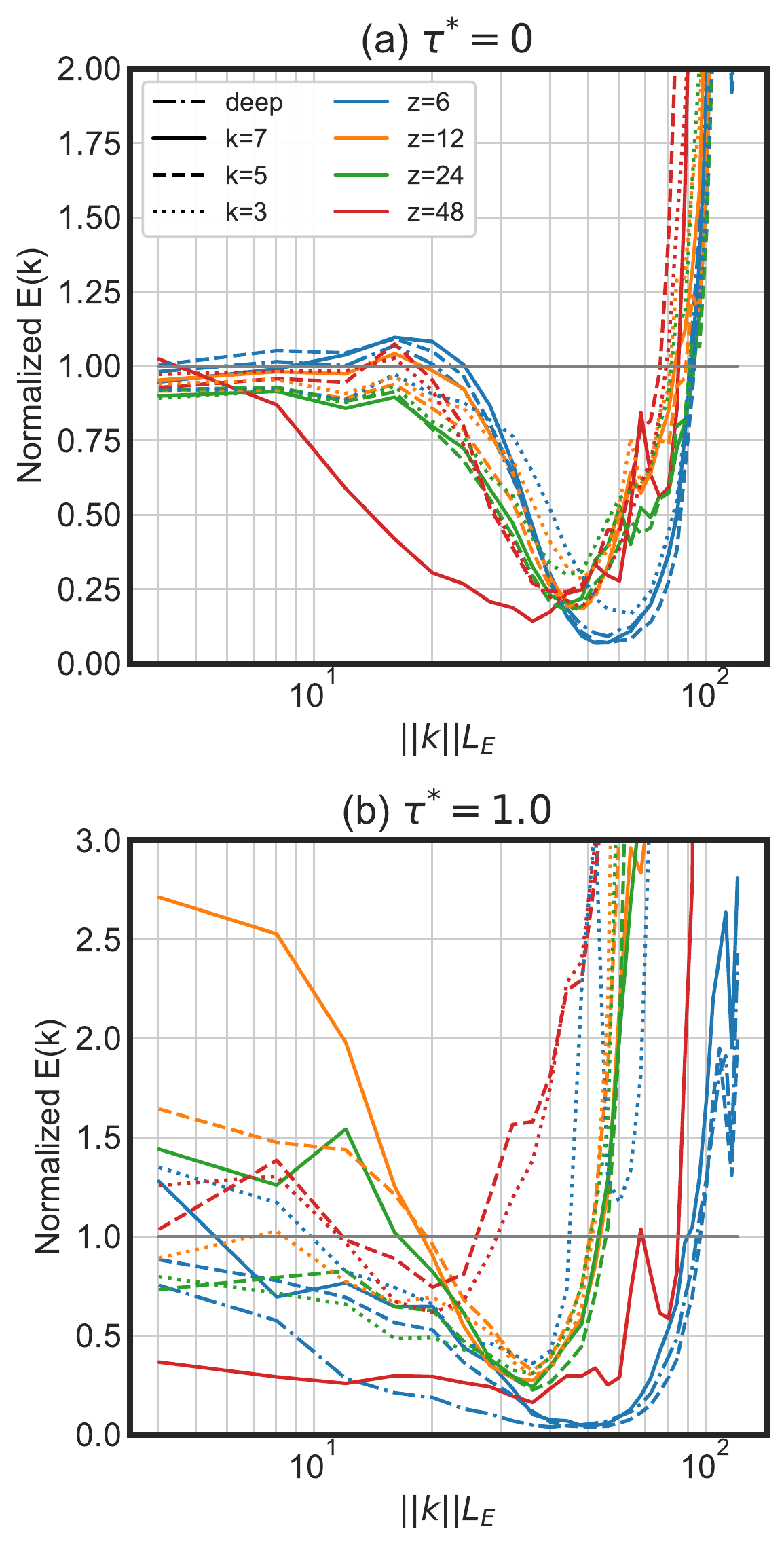}
  \caption{\small{Both plots show the energy of the predicted flow as a function of non-dimensionalized wavenumber, normalized to the energy of the true DNS data. (a) A snapshot of the energy spectra for all models corresponding to the initial time step $\tau^*=0$. Only the encoder/decoder networks impact the solution. (b) The energy spectra at the final time step $\tau^*=1$. For models with large receptive fields and latent spaces, low wavenumbers show better agreement with DNS, which are sufficient for many flow of engineering interest. Models with high compression in the latent and dynamics representations produce less realistic results with greater instability.} }
  \label{fig:spec}
\end{figure}

Next, kinetic energy powerspectra normalized by the true DNS spectra are shown in figure \ref{fig:spec} for the model predictions. In panel \ref{fig:spec}a, we show spectra from fields of the neural network model at the initial condition. The wavenumber $x-$axis is non-dimensionalized by $L_E$, given by:
\begin{equation}
L_E = \frac{e^{3/2}}{\epsilon} \quad .
\end{equation}
By comparing the DNS and neural network initial condition, denoted $\tau^\ast=0$, we first note the impact of the dimensionality reduction in removing smallest scale velocity fluctuations, whereby we observe a rapid attenuation of the predicted power spectra near $||k||L_E=20$. Again, $k7z48$ appears to be an outlier. We note that at this initial condition, the NeuralODE subnetwork has zero impact on the power spectra such that sources of error are constrained to the encoder/decoder subnetworks. With respect to the DNS solutions, we observe the neural network predictions to be accurate at small wavenumbers approximately less than 20, consistent with alternative reduced-order solution techniques such as large-eddy simulation.  

Fig \ref{fig:spec}b shows model spectra at $\tau^\ast=1$, where the DNS powerspectra are almost identical to those of the initial condition, owing to the forcing procedure applied to the system. Here, we see the effect of the temporal evolution of the models, which now differentiate themselves significantly at all wavenumbers. In general, the $z=6$ models are superior in accuracy, in the low wavenumber region below 20 and even in the highest wavenumber region, greater than 100.
We remark that the wavenumber regime where small scale fluctuations begin to attenuate at $\tau=0$ in the neural network model is strongly correlated with the wavenumber regime where deviations develop at $\tau=1$, indicating that the NeuralODE subnetwork is able to accurately model dynamics for the flow scales which are resolved by the encoder/decoder subnetworks.


We show the temporal velocity autocorrelations of the models, given as:
\begin{equation}
    \rho_{ii}(\tau^*)=\frac{\langle \mathbf{u}_i(0) \mathbf{u}_i(\tau^*)\rangle}{\langle \mathbf{u}_i(0)^2\rangle} \quad ,
\end{equation}
again normalized to the true DNS data, in figure \ref{fig:temporal}b. The models begin to diverge at $\tau^*=0.1$ with the three best performing networks, $k7z6$, $k5z6$, and $deep$, once again separating themselves from the rest. We also see the greater accuracy of models with larger NODE kernels by comparing autocorrelations within a family of models with the same compression ratio. $k7z48$ is an exception here as well. Figure \ref{fig:temporal}c presents the TKE-normalized mean-squared error over time. The 4 $z=6$ models exhibit the lowest error of all tested.

Real turbulent flows exhibit certain universal small-scale structures \cite{Elsinga2010UniversalAO}, such as a preference for vorticity alignment with the intermediate strain-rate eigenvalue. The flow topology can be described using the $\mathrm{Q-R}$ plane, which represent the second and third velocity gradient tensor invariants respectively. Isolines of probability in the $\mathrm{Q-R}$ plane, expressing intimate features of the turbulent flow geometry, have a nontrivial shape documented in the literature. Different parts of the $\mathrm{Q-R}$ plane are associated with different structures of the flow. Thus, the lower right corner (negative $Q$ and $R$), which has higher probability than other regions, corresponds to a pancake type of structure (two expanding directions, one contracting) with the direction of rotation (vorticity) aligned with the second eigenvector of the stress. We additionally perform coarse $\mathrm{Q-R}$ tests \cite{Chertkov1999}, which provide a description of the flow topology at varying length scales. The scale is described with the radius of the coarse-graining filter given in terms of the Kolmogorov length scale $\eta$. The tear-drop shape of the probability isolines becomes more prominent with decrease of the coarse-graining scale.

\begin{figure}
  \centering
  \includegraphics[width=\linewidth]{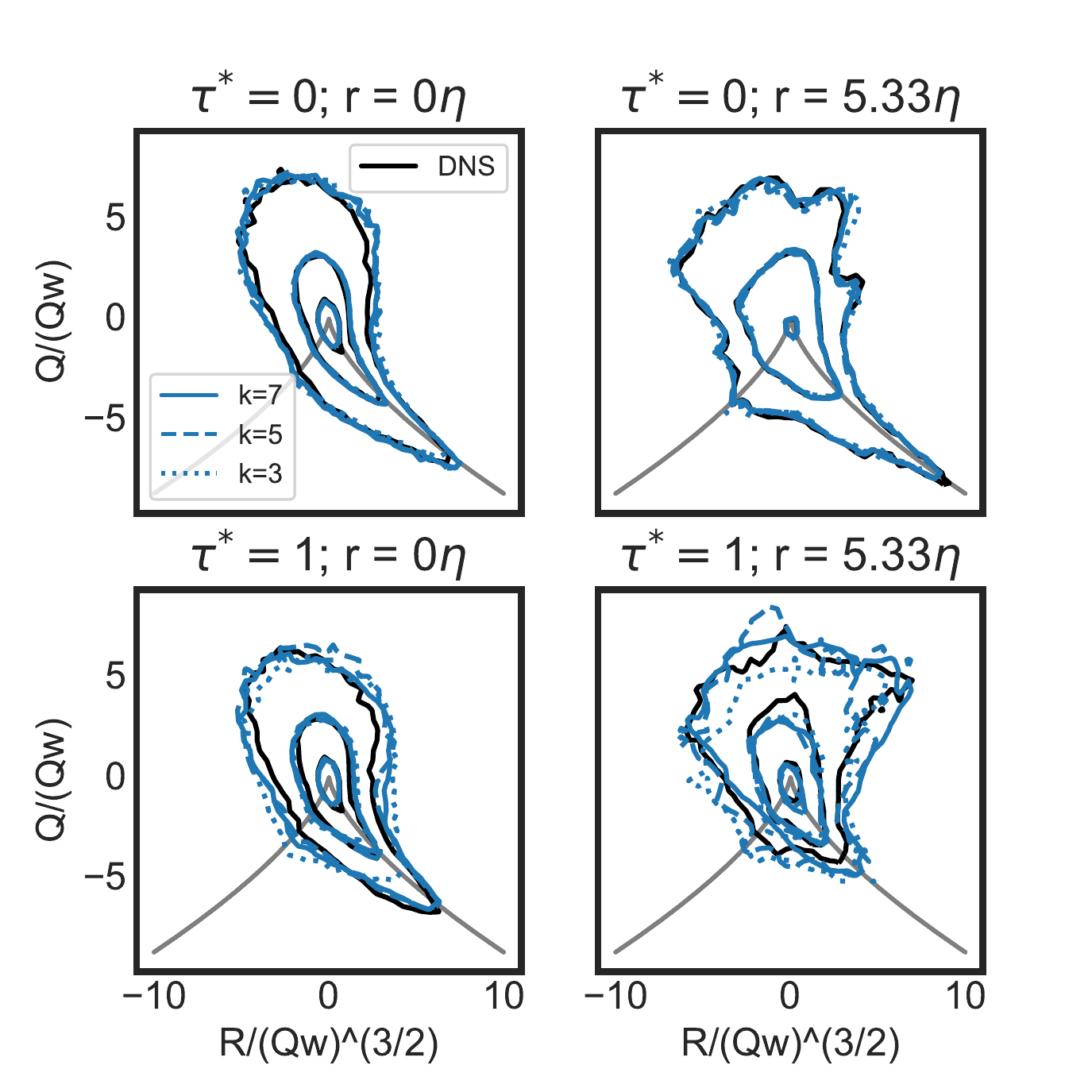}

  \caption{\small{$\mathrm{Q-R}$ planes of the flow shown at two length scales and two temporal snapshots for models with varying $k$. $z=6$ for all models shown. Rows correspond to different time steps, $\tau^*=0$ and $1.0$, and columns indicate the coarse-grained scales given in terms of the Kolmogorov microscale $\eta$.}}
  \label{fig:qr_k}

\end{figure}

\begin{figure}
  \centering
  \includegraphics[width=\linewidth]{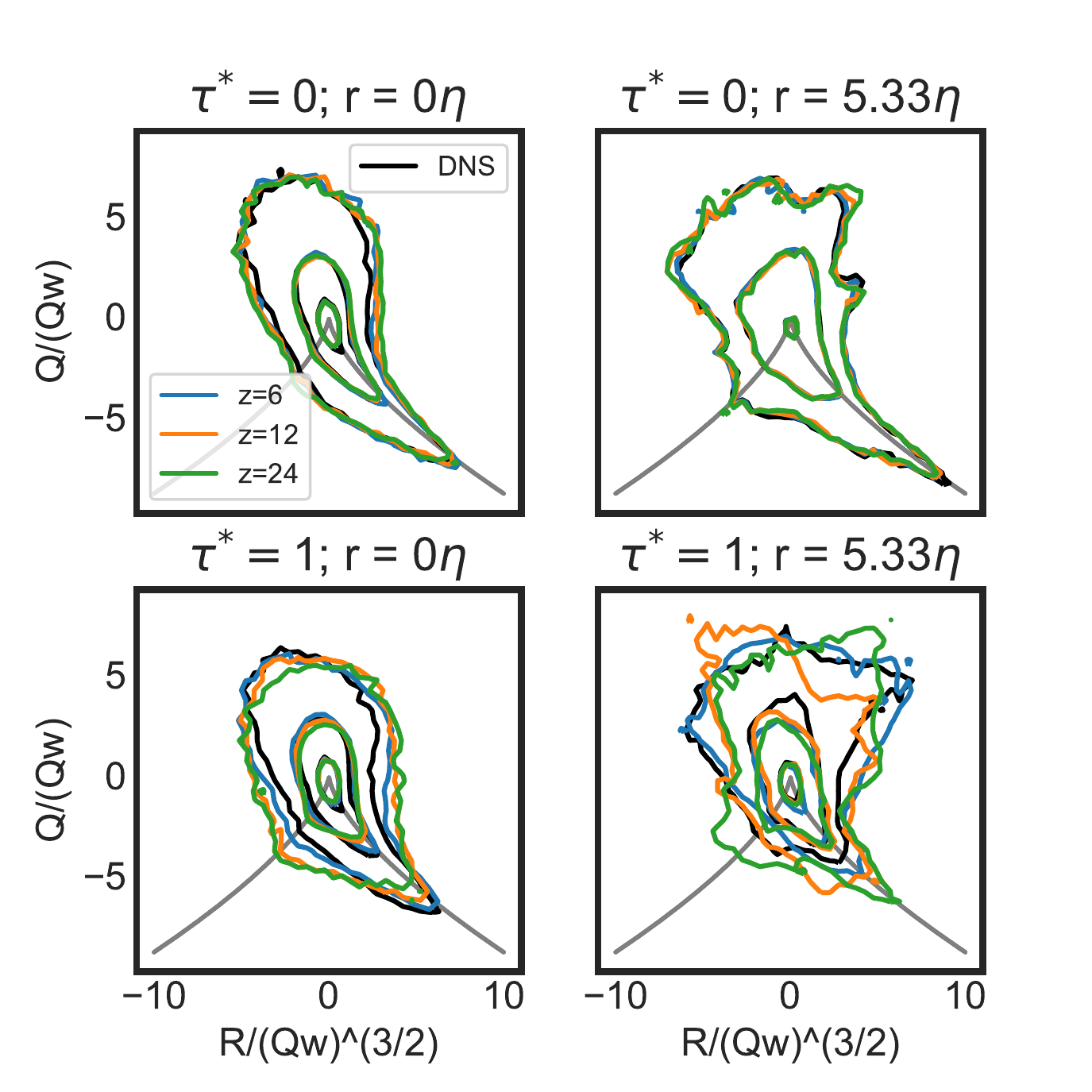}

  \caption{\small{$\mathrm{Q-R}$ planes of the flow shown at two length scales and two temporal snapshots for models with varying $z$. $k=7$ for all models shown. Rows correspond to different time steps, $\tau^*=0$ and $1.0$, and columns indicate the coarse-grained scales given in terms of the Kolmogorov microscale $\eta$.}}
  \label{fig:qr_z}

\end{figure}

Figures \ref{fig:qr_k} and \ref{fig:qr_z} compare the various network prediction $\mathrm{Q-R}$ distributions with the ground truth DNS distributions. Figure \ref{fig:qr_k} demonstrates the effect of NODE kernel size using the outputs from the $k7z6$, $k5z6$, and $k3z6$ models. The 4 panels provide insight into the PDFs at different coarse-graining scales and temporal snapshots. The rows correspond to the initial and final snapshots, $\tau^*=0$ and $\tau^*=1$, while the columns represent length scales $r=0\eta$, where no filter is used, and $r=5.33\eta$, $1/4$ of the length scale $L$ of the domain. As expected, at $\tau^*=0$, where the NODE effect is null, all three models perform almost identically. It is clear that at both length scales the flow morphology is well reconstructed by the encoder/decoder subsystem. At $\tau^*=1$, we start to see the differences between models. With no coarse-graining applied, there is a trend towards more symmetric PDFs as the kernel size is reduced. A symmetric distribution suggests the presence of random noise in the prediction that strays from the true statistical trends found in real turbulence. At $r=5.33\eta$, the teardrop shape of the DNS data is far less pronounced and there is greater variation between models. However, $k7z6$ best estimates the underlying true distribution.

Figure \ref{fig:qr_z} demonstrates the effect of compression ratio using the outputs from the $k7z6$, $k7z12$, and $k7z24$ models. At $\tau^*=0$, perhaps counterintuitively, we see that the morphology is well reconstructed at both length scales despite increases in compression ratio. All models perform similarly, and it appears the compression ratios tested have little bearing on the predicted morphology when no dynamics are involved. In contrast, at $\tau^*=1$ and $r=0\eta$, the trend towards more symmetric distributions as the compression ratio is increased is observed again. $k7z6$ also has the best large-scale structure conformity at $r=5.33\eta$ when compared to the other models with larger $z$.

\begin{figure}
  \centering
  \includegraphics[width=0.9\linewidth]{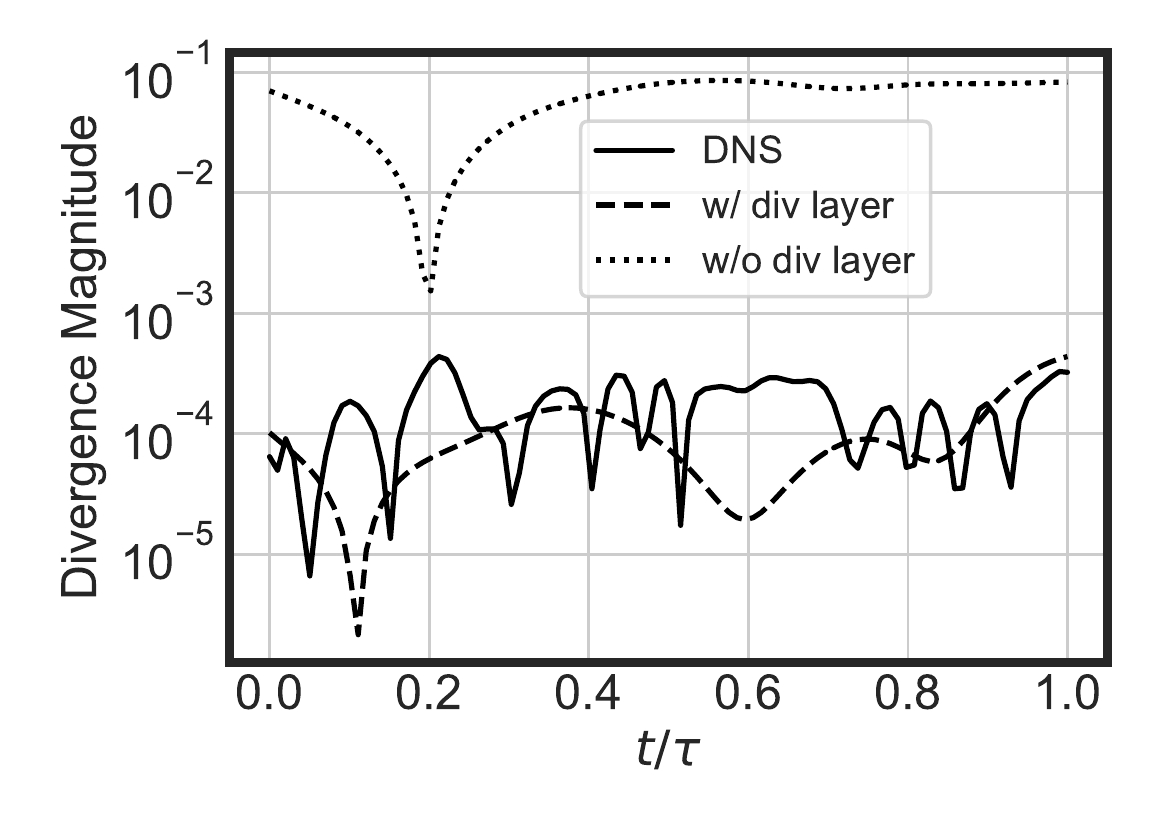}

  \caption{\small{The model uses a divergence-constraining layer to induce physical solution fields. The magnitude of the divergence is plotted as a function of time for three cases: the true divergence, the prediction from the model without any constraint layer, and the prediction from the same model with the divergence constraint. The difference between the predictions is significant, and it is clear the constrained model performs close to the expected output.}}
  \label{fig:div}

\end{figure}

Additionally, figure \ref{fig:div} shows the magnitude of divergence in the velocity field over one integral timescale for two models. The impact of the spectral projection step is significant in reducing the output divergence to magnitudes comparable to the true field. The formulation of the layer as a constrained optimization problem minimizing the deviations between $\mathbf{U}'$ and $\mathbf{\hat{U}}$, from eq. \ref{eq:div_opt}, implies inputs $\mathbf{u}'$ and outputs $\mathbf{\hat{u}}$ are highly comparable. The average correlation coefficient between $\mathbf{u}'$ and $\mathbf{\hat{u}}$ is 0.9986. This lends additional credence to the idea that the latent NeuralODE is learning the majority of the turbulent dynamics, and this step acts as a simple projection to correct the velocity field. The strong correlation also means training dynamics are not significantly affected by using the projection step, and MSE and other evaluation metrics are comparable between the models.

\section{Conclusions}
Encouraged by broader adoption of deep neural network architectures for reduced order modeling of physics problems, we explore a model design paradigm which attempts to reduce ambiguity in the parameterization of model error.   
We have approached this objective by considering an idealized problem setting of the incompressible Navier Stokes equations with spatio-temporal homogeneity, a system with strong nonlinearity and solution complexity.
In addition to implementing hard physical constraints, such as the implicit preservation of incompressibility, we have attempted to explicitly decouple individual modeling demands into discrete and interpretable subnetworks. Firstly utilizing convolutional encoders and decoders to reduce velocity field complexity, and also utilizing a NeuralODE architecture to model the dynamics imposed by the Navier Stokes equations.

We assess the methodology by analyzing forecasts from this scheme with varying model parameters and comparing them to ground truth DNS data. While qualitative comparisons to labeled data may be sufficient for non-scientific applications, we are interested in a rigorous study of the turbulent physics observed in the model’s outputs. Statistical metrics like energy spectra, $\mathrm{Q-R}$ plane PDFs, and autocorrelations are used to illustrate the quality of predictions. 

Across the board, we see this surrogate approach’s inclination to prefer preserving the majority energy-containing large eddies in the flow while filtering the smaller length scale fluctuations. To an extent, decreasing the latent space size still allows for conservation of energy at the large scales, but we see increasingly larger deviations in the temporal correlations and small-scale performance. In particular, there is a shift towards more symmetric distributions in the $\mathrm{Q-R}$ plane, indicating that the network is predicting noise instead of correlated statistics. The kernel size of layers in the dynamical CNN also imparts an effect on physical interpretations of the results. The smallest kernel network tested showed non-stationarity and energy loss even at the larger scales from the spectra, as well as divergence in the flow morphology at the small-, and to a lesser extent the medium- and large-scales. At the limit of our computational capabilities, model $k7z6$, we find excellent agreement with the ground truth. Some of the energy at the smaller length scales is lost, also reflected in the TKE, but because the smaller eddies contain less energy, over 90\% of the total energy in the flow is conserved. The divergence-free velocity fields from the forecasts imply mass in the system is conserved as well.

Building on previous research in scientific ML and turbulence modeling with ML, this work further demonstrates the effectiveness of data-driven models to be useful surrogates for approximating fluid dynamics. Data-driven convolutional NeuralODEs are multiple orders of magnitude faster as surrogates than traditional simulation methods and still capture important details of the flow, which supports use-cases where rapid predictions are essential, such as in design or parameter-space exploration applications. The addition of physics-grounded learning biases in these models is especially pragmatic for constraining the model and producing more physical outputs. We have also taken steps towards parameterizing the quality of such convolutional NeuralODE architectures, particularly for fluid dynamics applications, paving an avenue for practitioners to approach real-world engineering or other flow problems with data-driven methods. As the field of machine learning and fluid dynamics rapidly develops, others have been pushing the bounds of data-driven physics-informed models, even incorporating traditional solver machinery into their architectures. We see the space of possible models as a spectrum, where there is a tradeoff in computational cost and accuracy as one incorporates increasingly more physics into the model. It will be vital moving forward for practitioners to strike the right balance for their specific application, and therefore in depth studies of models along this spectrum will be crucial to the community.

\section{Acknowledgements}
This material is based upon work supported by the National Science Foundation Graduate Research Fellowship under Grant No. DGE 1745016 awarded to VS.  The authors from CMU acknowledge the support from the Technologies for Safe and Efficient Transportation University Transportation Center, and Mobility21, A United States Department of Transportation National University Transportation Center. This work was supported in part by Oracle Cloud credits and related resources provided by the Oracle for Research program. The authors further acknowledge the support of NVIDIA via The ICEnet Consortium, for providing necessary GPU compute.  A.T.M and G.D.P. have been supported by the LDRD (Laboratory Directed Research and Development) program at LANL under project 20190059DR.

This work has been co-authored by employee(s) of Triad National Security, LLC which operates Los Alamos National Laboratory (LANL) under Contract No. 89233218CNA000001 with the U.S. Department of Energy/National Nuclear Security Administration.  
\bibliography{bib}
\end{document}